\documentclass[sigconf]{acmart}

\usepackage{hyperref}
\usepackage{url}
\usepackage{xcolor}
\usepackage{enumerate}
\usepackage{textcomp}
\usepackage{mathrsfs}
\usepackage{amsthm}
\usepackage{bbm}
\usepackage{epigraph}
\usepackage{etoolbox}
\usepackage{thmtools, thm-restate}
\usepackage{proof-at-the-end}
\newtheorem{example}{Example}

\usepackage{geometry}
\usepackage{algorithm}
\usepackage{algpseudocode}
\usepackage{float}

\usepackage{listings}

\usepackage{xparse}

\NewDocumentCommand{\codeword}{v}{%
\texttt{\textcolor{blue}{#1}}%
}
\NewDocumentCommand{\bill}{v}{%
\texttt{\textcolor{green}{#1}}%
}
\NewDocumentCommand{\yt}{v}{%
\texttt{\textcolor{blue}{#1}}%
}

\def\normal{\mathcal{N}}

\def\E{\mathbb{E}}

\def\Pr{\mathbb{P}}

\def\1{\mathbf{1}}

\newcommand{\Ac}{\mathcal{A}}

\newcommand{\Vc}{\mathcal{V}}

\newcommand{\Lc}{\mathcal{L}}
\newcommand{\Oc}{\mathcal{O}}

\usepackage{hyperref}

\definecolor{codegreen}{rgb}{0,0.6,0}
\definecolor{codegray}{rgb}{0.5,0.5,0.5}
\definecolor{codepurple}{rgb}{0.58,0,0.82}
\definecolor{backcolour}{rgb}{0.95,0.95,0.92}

\lstdefinestyle{mystyle}{
    backgroundcolor=\color{backcolour},   
    commentstyle=\color{codegreen},
    keywordstyle=\color{magenta},
    numberstyle=\tiny\color{codegray},
    stringstyle=\color{codepurple},
    basicstyle=\ttfamily\footnotesize,
    breakatwhitespace=false,         
    breaklines=true,                 
    captionpos=b,                    
    keepspaces=true,                 
    numbers=left,                    
    numbersep=5pt,                  
    showspaces=false,                
    showstringspaces=false,
    showtabs=false,                  
    tabsize=2
}
\usepackage[mathscr]{euscript}
\hypersetup{
    colorlinks=true,
    linkcolor=blue,
    filecolor=magenta,      
    urlcolor=purple,
}
\usepackage{xcolor}

\newcount\Comments
\newcommand{\kibitz}[2]{\ifnum\Comments=1{\textcolor{#1}{\textsf{\footnotesize #2}}}\fi}
\usepackage{color}
\definecolor{darkred}{rgb}{0.7,0,0}
\definecolor{darkgreen}{rgb}{0.0,0.5,0.0}
\definecolor{darkblue}{rgb}{0.0,0.0,0.5}
\definecolor{teal}{rgb}{0.0,0.5,0.5}

\AtBeginDocument{%
  }

\setcopyright{acmlicensed}
\copyrightyear{2024}
\acmYear{2024}
\acmDOI{XXXXXXX.XXXXXXX}

\acmConference[WebConf '25]{Make sure to enter the correct
  conference title from your rights confirmation email}{April 28--May 02,
  2025}{Sydney, Australia}
\acmISBN{978-1-4503-XXXX-X/18/06}

\begin{document}

\title{Epinet for Content Cold Start}

\author{Hong Jun Jeon}
\email{hjjeon@stanford.edu}
\affiliation{%
  \institution{Stanford University}
  \city{Stanford}
  \state{California}
  \country{USA}
}

\author{Songbin Liu}
\affiliation{%
  \institution{Meta}
  \city{New York}
  \state{New York}
  \country{USA}}
\email{songbin@meta.com}

\author{Yuantong Li}
\affiliation{%
  \institution{Meta}
  \city{New York}
  \state{New York}
  \country{USA}}
\email{yuantongli@meta.com}

\author{Jie Lyu}
\affiliation{%
  \institution{Meta}
  \city{New York}
  \state{New York}
  \country{USA}}
\email{jlyu@meta.com}

\author{Hunter Song}
\affiliation{%
  \institution{Meta}
  \city{New York}
  \state{New York}
  \country{USA}}
\email{huntersong21@meta.com}

\author{Ji Liu}
\affiliation{%
  \institution{Meta}
  \city{Bellevue}
  \state{Washington}
  \country{USA}}
\email{ji.liu.uwisc@gmail.com}

\author{Peng Wu}
\affiliation{%
  \institution{Meta}
  \city{New York}
  \state{New York}
  \country{USA}}
\email{wupeng@meta.com}

\author{Zheqing Zhu}
\affiliation{%
  \institution{Meta}
  \city{Bellevue}
  \state{Washington}
  \country{USA}}
\email{billzhu@meta.com}


\begin{abstract}
    The exploding popularity of online content and its user base poses an evermore challenging matching problem for modern recommendation systems.  Unlike other frontiers of machine learning such as natural language, recommendation systems are responsible for collecting their own data.  Simply exploiting current knowledge can lead to pernicious feedback loops but naive exploration can detract from user experience and lead to reduced engagement.  This exploration-exploitation trade-off is exemplified in the classic multi-armed bandit problem for which algorithms such as upper confidence bounds (UCB) and Thompson sampling (TS) demonstrate effective performance.  However, there have been many challenges to scaling these approaches to settings which do not exhibit a conjugate prior structure.  Recent scalable approaches to uncertainty quantification via epinets \cite{osband2023epistemic} have enabled efficient approximations of Thompson sampling even when the learning model is a complex neural network.  In this paper, we demonstrate the first application of epinets to an online recommendation system.  Our experiments demonstrate improvements in both user traffic and engagement efficiency on the Facebook Reels online video platform.
\end{abstract}




\maketitle

\section{Introduction}
With the exponential growth of online content, the matching problem of user and content becomes evermore challenging.  This describes the task of modern recommendation systems and fortunately, it is \emph{far} from a needle in the haystack problem: \emph{past} user engagement is often indicative of \emph{future} engagement with a piece of content.  Furthermore, new content often still exhibits similarities to that of the past.  As a result, intelligent algorithm design ought to enable systems to reliably recommend content which each user individually finds engaging.

While other areas of machine learning such as natural language have captured the attention of academics and industry practitioners alike, the recommendation system problem remains a formidable frontier for algorithmic advancements.  Unlike natural language, the recommendation system problem falls squarely in the bandit learning or (more generally) reinforcement learning frameworks.  Bandit and RL problems differ from supervised learning in the crucial detail that the algorithm is tasked with collecting its \emph{own} data.  For recommendation systems, this entails that the algorithm must decide which content to recommend to each user and will only observe labels for the user/content pairs which it proposes.  While simple to describe, bandit problems exhibit a deep challenge known as the ``exploration vs exploitation trade-off''.  At any moment in time, an effective algorithm will have to balance the importance of accumulating short-term reward by \emph{exploiting} information that it already has collected, versus \emph{exploring} the unknown to acquire new information.

While there are known algorithms which optimally solve the multi-armed bandit problem \citep{gittins1979bandit}, the solutions are computationally intractable in all but the simplest of problem instances.  As a result, the past decade has seen an explosion of interest in \emph{efficient} algorithms which exhibit an effective trade-off between exploration and exploitation. 
Theoretical analyses of even the simplest multi-armed bandit problem settings have demonstrated that naive exploration approaches such as $\epsilon$-greedy and Boltzmann exploration result in $O(T)$ regret, where $T$ is the number of environment interactions.  As regret measures the difference between the reward of the algorithm's action and the optimal reward, linear regret implies that the agent will \emph{not} identify the optimal arm even with infinite environment interactions.  Meanwhile, methods such as Thompson Sampling (TS) and Upper Confidence Bounds (UCB) can achieve $O(\sqrt{T})$ regret, implying that these algorithms reliably identify the best bandit arm with repeated environment interaction.  The key insight is that algorithms which effectively leverage their \emph{epistemic uncertainty} (posterior distribution for TS and confidence interval for UCB) are able to provide fruitful strategies to combat the exploration-exploitation trade-off.  While the multi-armed bandit problem is an idealized setting which deviates from the realities of modern recommendation systems, it elucidates the unique benefits which are afforded by algorithms which effectively represent and leverage uncertainty.

The challenge of adapting UCB and TS to modern recommendation systems lies in tractably representing uncertainty for complex models involving neural networks.  While posterior distributions exhibit efficient updates in the presence of nice conjugate priors, this is certainly not the case for likelihoods involving neural networks.  However several lines of work have attempted to establish \emph{approximate} methods of posterior sampling.  Among the early works is stochastic gradient langevin dynamics \cite{welling2011bayesian} which adds noise to the stochastic gradient updates.  Upon arriving at a local minima the noise produces samples surrounding the posterior mode.  However, the algorithm is computationally onerous for online methods as it requires optimization to convergence at every timestep to produce samples from the posterior.  There have also been various works in the Bayesian Neural Network literature \citep{der2009aleatory, kendall2017uncertainties} which represent uncertainty via updating a posterior distribution over parameters.  However these methods are again computationally onerous and result in inadequate posterior approximation (factorized distributions on network weights are a poor characterization of model uncertainty).  Other works have proposed MC Dropout as a posterior sampling approximation but \cite{osband2016risk, hron2017variational} have demonstrated that the quality of the posterior approximation can be very poor.  Deep ensembles have been a reliable method to approximate posterior sampling via sampling from the ensemble's particles but suffer from the fact that compute requirements grow linearly in the size of the ensemble \citep{deep_ensemble, lu2017ensemble}.  Most recently, \cite{osband2023epistemic} have proposed \emph{epinet}, a computationally tractable approach which attains ensemble level performance with a fraction of the computational requirements.  \cite{zhu2023scalable} have applied epinet to an offline recommendation system problem but whether this method can produce meaningful improvements in an online production setting is still an open question.

In this paper, we provide the first online experimental deployment of epinet in a production recommendation system.  Specifically, we use a Thompson Sampling algorithm with posterior samples provided by epinet to produce recommendations in the cold start retrieval phase of Facebook's Reels recommendation.  Content cold start involves making recommendations for new content which does not have many impressions yet.  As a result it is a setting which requires effective navigation of the explore/exploit tradeoff.  Our experiments demonstrate improvements in overall traffic and engagement efficiency metrics such as like rate and video view completion rate across the cold start content corpus.

\section{Related Works}

In recommendation systems broadly and especially in content cold start, researchers have leveraged the multi-armed bandit formulation to design algorithms which effectively trade-off novel exploration of the content corpus and exploitation of information which has already been gathered \citep{caron2013mixing, nguyen2014using, felicio2017multi}.  
However, as the size of the content corpus and user base grows, the assumptions of statelessness and independent arms becomes increasingly unreasonable.  As a result, many works have come use probabilistic models which exhibit generalization across arms and contextualization which reflect the varying preferences of different users \citep{10.1145/2736277.2741104, 8023425, yue2011linear, 10.1145/3297280.3297440}.  However, even with these problem formulations, the design of algorithms which effectively solve these problems has been an active area of research.

As aforementioned, the most popular algorithms which exhibit nontrivial exploration are UCB \citep{agrawal1995sample, auer2002finite} and TS \citep{thompson1933likelihood}.  However, these algorithms were developed for the standard multi-armed bandit problem involving independent arms and no notion of state.  As a result, translating these algorithmic solutions to problem settings involving contextualization and generalization has been a pressing challenge in recent years.
In particular, with the dramatic increases in capabilities afforded by machine learning with deep neural networks, many methods have attempted to adapt the ideas from UCB and TS to be amenable to computation with neural networks
\citep{abeille2017linear,dwaracherla2020hypermodels,li2010contextual,xu2020neural,zhang2020neural,zhou2020neural,osband2015bootstrappedthompsonsamplingdeep}.  However, the major drawback with the above methods is that they are impractical when computational costs are factored in.  Notably, with the rise of recommendation systems which leverage immense transformer models to inform predictions \citep{zhai2024actions,qiu2021u,hu2024enhancing,wu2021empowering}, each flop of compute which is expended on resources outside of model or dataset size comes at a steep cost \citep{kaplan2020scaling, hoffmann2022training}.  To ameliorate these concerns surrounding computational resources, \citet{osband2023epistemic} have proposed \emph{epinet}, a computationally tractable approach which attains ensemble level performance with a fraction of the computational requirements.  \cite{zhu2023scalable} have applied epinet to an offline recommendation system problem but our work marks the first application of this technology in an online production recommendation system setting.

\section{Problem Formulation}

We formalize the recommendation system problem as a non-stationary contextual bandit problem.  
Concretely, the non-stationary contextual bandit problem can be identified by a tuple $$(\Oc, \Ac, (\Vc_t)_{t\in\mathbb{N}}, (\psi_t)_{t\in\mathbb{N}}, \rho),$$
where $\Oc$ denotes the observation set, $\Ac$ the action set, $(\Vc_{t})_{t\in\mathbb{N}}$ denotes the random process which dictates the non-stationarity, $(\psi_t)_{t\in\mathbb{N}}$ denotes the random process which dictates the \emph{contexts}, and $\rho$ denotes the probability associated with an observation $o\in \Oc$.  In the recommendation system problem, a context $\theta_t$ represents the raw features of a particular \emph{user}, while $\Vc_{t}$ represents the raw features of available \emph{content} to recommend to said user.  We dive into the details of each component below.
\begin{enumerate}
    \item Action space: We take $\Ac$ to be the fixed set $\{1, 2, \ldots, N\}$, where $N$ denotes the number of items to select from.  At the retrieval stage of recommendation, the algorithm can propose $M$ items per timestep.  Therefore, for all $t$, we let random variable $A_t:\Omega\mapsto\Ac^M$ denote the action taken by our algorithm at time $t$.
    \item Non-stationary item pool: We take $(V_t)_{t\in\mathbb{N}}$ to be the random process which represents the changing item pool.  For all $t$, $V_t = (\phi_{t,1}, \phi_{t,2}, \ldots, \phi_{t, N}) \in \Vc$ where $\phi_{t,i}$ denotes the raw features associated with item $i$ at time $t$ and $\Vc$ denotes the range of $V_t$.  Since we are focused on content cold start, the item pool is constantly being refreshed by adding newly created content and removing matured content. 
    \item Context: We take $(\psi_t)_{t\in\mathbb{N}}$ to be the random process which represents the user context at each timestep.  For all $t$, $\psi_t\in\Psi$ consists of the raw features associated with the user for which we provide recommendations.
    \item Observation Space: We take $\Oc$ to be set of observations about a user and a recommended item.  secommendation systems often leverage multiple signals including binary signals such as whether the user liked or shared the item, and real-valued signals such as the proportion of the video which was completed.  However, to make actions for the following timestep, the algorithm must have access to the subsequent item pool $V_{t+1}$ and context $\psi_{t+1}$.  Therefore, for all $t$, we let random variable $O_{t+1}: \Omega\mapsto\Oc^M \times \Vc \times \Psi$ denote the observation associated with action $A_t$, the subsequent item pool $V_{t+1}$, and the subsequent context $\psi_{t+1}$.  $M$ again denotes the number of items that were proposed in action $A_t$.
    \item Observation probability: We let $\rho(\cdot|O_t, A_t) = \Pr(O_{t+1}\in\cdot|O_t, A_t)$ denote the probability of a subsequent observation conditioned on the previous observation and action.  Recall that $O_{t}$ also contains the item pool $V_t$ and the user context $\psi_t$.  Generalization across users and items with similar features is captured by $\rho$.
\end{enumerate}

We assume that the algorithm designer has an concrete objective in mind.  This objective is represented by a (known) reward function $\mathcal{R}: \mathcal{O}\mapsto \Re_+$ which maps an observation to a scalar objective value.  In practice, this is a suitable weighting of the various labels (like, share, etc).  We let random variable $R_{t+1}$ denote the sum of rewards of all $M$ labels observed in $O_{t+1}$.  The objective is to maximize the average long-term reward:
$$\frac{1}{T}\sum_{t=0}^{T-1}\E\left[R_{t+1}\right].$$

\section{Exploration for Content Cold Start}

Cold start content consists of videos which have been shown to fewer than 10000 users.  Note that even if the video is shown to a user, this does not necessarily mean that the user meaningfully \emph{engaged} with it.  In fact it is rather the contrary as for instance, only $1\%$ of recommended content results in a like from the user.  With this sparse feedback signal, the ability to learn efficiently becomes an imperative.  Algorithms which explore ineffectively will not only provide poor recommendations to the user, but will also consume valuable bandwidth which could have been allocated to better learning about the user's preferences.

The multi-armed bandit problem, despite its simplicity, assesses a core capability for an intelligent agent: The ability to balance immediate reward with the acquisition of new information (exploration/exploitation tradeoff).  To effectively direct exploration of the content corpus, an algorithm must \emph{know what it doesn't know}.  In the Bayesian framework, uncertainty is modeled using the tools of probability and represented via a \emph{posterior distribution}.  As the algorithm observes more data, this distribution will concentrate, reflecting greater \emph{certainty}.  This uncertainty which is reducible by observing more data is typically referred to as \emph{epistemic uncertainty}.  Common algorithms such as Thompson sampling and upper confidence bounds represent epistemic uncertainty via a posterior distribution and a confidence interval respectively.

However many production recommendation systems keep only a \emph{point estimate}.  A point estimate corresponds to a distribution in which all probability mass is placed on a singleton.  This does not effectively model epistemic uncertainty since the algorithm is equally (and absolutely) certain across all time.  This is especially problematic for cold start content since it consist of items with little to no impression data.  The point estimate model's score for such content is essentially uninformed.  If it happens to \emph{underestimate} the value of the content, it will never be proposed to the user.  This could further exacerbate the issue referred to as \emph{popularity bias}, the tendency of recommendation systems to simply \emph{exploit} based off of existing data and forego exploration.  Exploration for recommendation systems can therefore be reduced to representing epistemic uncertainty for complex modern systems involving deep neural networks.

\subsection{Thompson Sampling}
In this paper, we focus on representing epistemic uncertainty with approximate posterior distributions.  Therefore, we limit our algorithmic focus to Thompson sampling, which samples from this posterior distribution to select actions.  In this section, we provide an abstract overview of Thompson sampling to frame our eventual algorithm.

For all $t$, Thompson sampling takes as input an approximate posterior distribution $P_t(\rho\in\cdot)$ and the previous observation $O_t$, and produces an action $A_t$.  Recall that $\rho$ is the probability measure which dictates the observations in our contextual bandit environment.  The following pseudocode abstractly depicts Thompson sampling:

\begin{algorithm}[h]
	\caption{Thompson Sampling} 
	\begin{algorithmic}[1]
            \State Initialize $P_0(\cdot)$
            \State Observe $O_0$
		\For {$t =0,1,2,\ldots$}
                \State Sample $\hat{\rho} \sim P_t(\cdot|O_t)$
                \State Select $M$ actions $a_1, \ldots, a_M \in \Ac$ which maximize $\E[R_{t+1}|A_t=a, \rho=\hat{\rho}]$
                \State $A_t \leftarrow (a_1, a_2, \ldots, a_M)$
                \State Observe $O_{t+1}$
                \State $P_{t+1} \leftarrow {\rm (Approximate)\ Bayesian\ Update}(P_t, A_t, O_{t+1})$
		\EndFor
	\end{algorithmic} 
\end{algorithm}

Therefore, approximate Thompson sampling methods differ in $1)$ how the specify the approximate prior distribution $P_0(\cdot)$ and $2)$ how they approximate the Bayesian update.  While simplifying assumptions such as independent beta prior/posterior distributions may enable exact Bayesian updating, they greatly suffer from the inability to account for information present in the user context $\psi_t$ and the item features $V_t$.

\section{Approximate Thompson Sampling with Epinet}

In this section we present our method which leverages epinet \citep{osband2023epistemic} to model epistemic uncertainty in an approximate Thompson sampling algorithm.  We begin with some background on \emph{epistemic neural networks}, a broad class of approximate posterior methods which encompass epinet.

\subsection{Epistemic Neural Networks}
 
\cite{NEURIPS2023_07fbde96} defined a class of neural networks known as \emph{epistemic neural networks} (ENNs).  Epistemic neural networks are specified by a parameterized function class $f$ and a reference distribution $P_Z$.  The output $f_\theta(X,z)$ depends both on the input $X$ and an epistemic index $z$ which is drawn from the reference distribution $P_Z$.  To produce a marginal prediction for an input $X$, an epistemic neural network \emph{integrates} over its epistemic indices: 
$$\hat{P}(Y) = \int_{z} f_{\theta}(X, z)dP_Z(z).$$
Many existing methods of uncertainty modelling for neural networks including BNNs, MC Dropout, and Deep Ensembles can be expressed as ENNs under a suitable reference distribution $P_Z$.  We instantiate a few below.

\begin{example}{\bf (Point Estimate)}
    Consider a parameterized neural network $g_\theta(\cdot): \Re^d\mapsto\Re$.  Let $P_Z(\cdot)$ be \emph{any} distribution.  Then $f_{\theta}(X,z) = g_\theta(X)$ is equivalent to a network which only keeps a point estimate.
\end{example}

In our experimentation, the baseline that we compare our algorithm to is the point estimate.  This is because it is the standard method in machine learning and recommendation systems: provide predictions/recommendations which are optimal according to the learned model.  While this is a reasonable approach in supervised learning settings in which the algorithm does not have control over the data that it observes, in a bandit/reinforcement learning problem, this exploitative behavior can result in feedback loops such as popularity bias in recommendation systems.

\begin{example}{\bf(MC Dropout)}
    Consider a parameterized neural network $g_\theta(\cdot): \Re^d\mapsto\Re$ with dropout applied to the input layer.  Let $P_Z(\cdot) = {\rm Uniform}\left( \{0,1\}^d \right)$.  Then, $f_\theta(X,z) = g_\theta(X\odot z)$ is equivalent to dropout.
\end{example}
We do not directly compare performance against MC dropout as it has been demonstrated theoretically and empirically to provide poor representations of epistemic uncertainty \cite{osband2016risk, osband2023epistemic}.  To see this, one need look no further than the fact that the variance of MC dropout does not go to $0$ even as the number of observed examples increase to $\infty$.  \citet{osband2016risk} identifies that this is because dropout approximates aleatoric risk as opposed to epistemic uncertainty.  This distinction is key as using the former to guide exploration is incoherent while using the latter can result in efficient exploration.

\begin{example}{\bf(Deep Ensembles)}
    Consider a deep ensemble of size $N$ comprised of parameterized models $g_{\theta_1}(\cdot), g_{\theta_2}(\cdot), \ldots, g_{\theta_N}(\cdot): \Re^d\mapsto\Re$.  Let $P_Z(\cdot) = {\rm Uniform}\left(\{1,\ldots, N\}\right)$.  Then, $f_{\theta}(X,z) = g_{\theta_{z}}(X)$ is equivalent to deep ensembles.
\end{example}

Deep ensembles have been shown to provide useful representations of epistemic uncertainty in bandit and reinforcement learning settings \cite{lu2017ensemble, qin2022analysis}.  They have even demonstrated good performance in offline recommendation system settings \cite{10.1145/3604915.3608855}.  However, their major drawback is their enormous computational overhead.  The computational resources scale linearly in the size of the ensemble.  Ensembles typically need on the order of $100$ particles to function effectively, a figure which is computationally infeasible when each base model is operating at the scale of Facebook's recommendation system.  As a result, we omit experimentation with ensembles in favor of Epinet, a method which promises ensemble-level performance without ensemble-level computational resources \cite{osband2023epistemic}.

\subsection{Epinet}

Recent work has demonstrated that epinet \cite{osband2023epistemic} can achieve performance comparable to deep ensembles with hundreds of particles with only a small fraction of the computational overhead.  We detail the design and implementation of epinet in this section.

To integrate epinet into Meta's existing cold start retrieval model, we create some mild modifications from the original outline of \cite{osband2023epistemic} and \cite{zhu2023scalable}.  The system consists of two neural networks: a user tower $g^{\rm (user)}_{\xi_u}$ with parameters $\xi_{u}$ and an item tower $g^{\rm (item)}_{\xi_i}$ with parameters $\xi_{i}$.  The system is trained on several supervision signals such as like, share, etc. Let $K$ denote the number of such signals.  The item tower takes the raw features $\phi_{t,a}$ of an item and outputs an embedding vector 
$$\underbrace{g^{\rm (item)}_{\xi_i}(\phi_{t,a})}_{\rm item\ embeddings} \in \Re^{d}.$$
Meanwhile, the user tower takes the raw features $\psi_{t}$ of the user context and outputs an embedding matrix $$\underbrace{g^{\rm (user)}_{\xi_u}(\psi_{t})}_{\rm user\ embeddings} \in \Re^{d\times K}.$$  
These user and item embeddings are fed into an \emph{overarch} model which includes the epinet.  The overarch consists of a base mlp and an epinet.  The base mlp $g_{\theta}$ consists of trainable parameters $\theta$ but does not depend on a sampled epistemic index $z$.  Meanwhile, the epinet $\sigma_\eta$ consists of trainable parameters and takes as input a sampled epistemic index $z$.  The base mlp and the epinet both take as input the concatenated user and item embeddings as well as in interaction term which consists of the concatenated elementwise product of the item embedding with the $K$ user embeddings.  Hence, for all $t$, the input is hence a vector $x_t \in \Re^{d(2K+1)}$. 
 The overarch output is as follows:
$$\underbrace{f_{\theta,\eta}(x_t,z_t)}_{\rm overarch}\ =\ \underbrace{g_{\theta}({\rm sg}[x_t])}_{\rm base\ mlp} + \underbrace{\sigma_{\eta}({\rm sg}[x_t], z_t)}_{\rm epinet},$$
where ${\rm sg}[\cdot]$ denotes a stop gradient.  The stop gradient has been observed to improve training stability \citep{osband2023epistemic}.

The additive form of epinet is motivated by two factors $1)$ functional uncertainty estimation and $2)$ efficient uncertainty estimation.  Functional uncertainty estimates emphasize that neural network uncertainty is \emph{not} over the model parameters, but rather the \emph{function} that the parameters induce.  This is an important distinction as many different arrangements of the parameters result in the \emph{same} function (appropriate permutations of the neurons for instance).  Efficient uncertainty estimation is possible if the epinet is smaller than the base model and the computation for evaluating several epistemic indices $z_1, \ldots, z_n$ can be batched.  For methods such as MC dropout, deep ensembles, and BNNs, either the prohibitive size of the models and/or the inability to perform batched computation makes uncertainty computation burdensome.  We now provide further architectural details of the epinet.

\subsection{Prior Networks}

In the literature, it is common to further divide the epinet into a learnable component and a fixed component referred to as a \emph{prior network} \citep{osband2023epistemic, zhu2023scalable}.  \citet{dwaracherla2022ensemblesuncertaintyestimationbenefits} have demonstrated that for deep ensembles, prior functions dramatically improve upon vanilla deep ensembles \cite{deep_ensemble} particularly in low-data regimes.  Concretely, if $x_t$ denotes the epinet input and $z_t$ the epistemic index, the epinet takes the form
$$\underbrace{\sigma_{\eta}(x_t, z_t)}_{\rm epinet} \ =\ \underbrace{\sigma_{\eta}^{L}(x_t, z_t)}_{\rm learnable} + \underbrace{\sigma^P(x_t, z_t)}_{\rm prior\ net}.$$
We assume that the epinet produces a scalar output and that $z_t \overset{iid}{\sim}\normal(0, I_{d_z})$.  While the embeddings are trained on several labels (like, share, etc.), for our experimentation, we limited training the epinet to only a single label.  We note that the prior network is \emph{not} trainable.  In both deep ensembles and epinet, these prior functions appear to be crucial algorithmic additions to ensure sufficient initial diversity within the represented distribution (variability in $z$).

The choice of architecture for the epinet is rather peculiar but it draws inspiration from \emph{linear bandits}.  The extension to neural networks provided in \citep{NEURIPS2023_07fbde96} is ad-hoc but they demonstrate good performance nonetheless.  Deriving more appropriate forms for softmax/sigmoidal outputs is a potential direction for future work.  The learnable network is a standard multi-layered perceptron (MLP) with Glorot initialization:
$$\sigma_\eta^{L}(x_t, z_t) \ =\ {\rm mlp}_{\eta}\left([x_t, z_t]\right)^\top z_t,$$
where ${\rm mlp}_{\eta}$ returns an output in $\Re^{d_z}$ and $[x_t, z_t]$ is a concatenation of $x_t$ and $z_t$.  Meanwhile, typical choices of the prior network include $\sigma^P$ sampled from the same architecture as $\sigma^L$ but with different parameter initialization.

\subsection{Model Training}

\begin{figure*}
    \centering
    \includegraphics[width=0.8\textwidth]{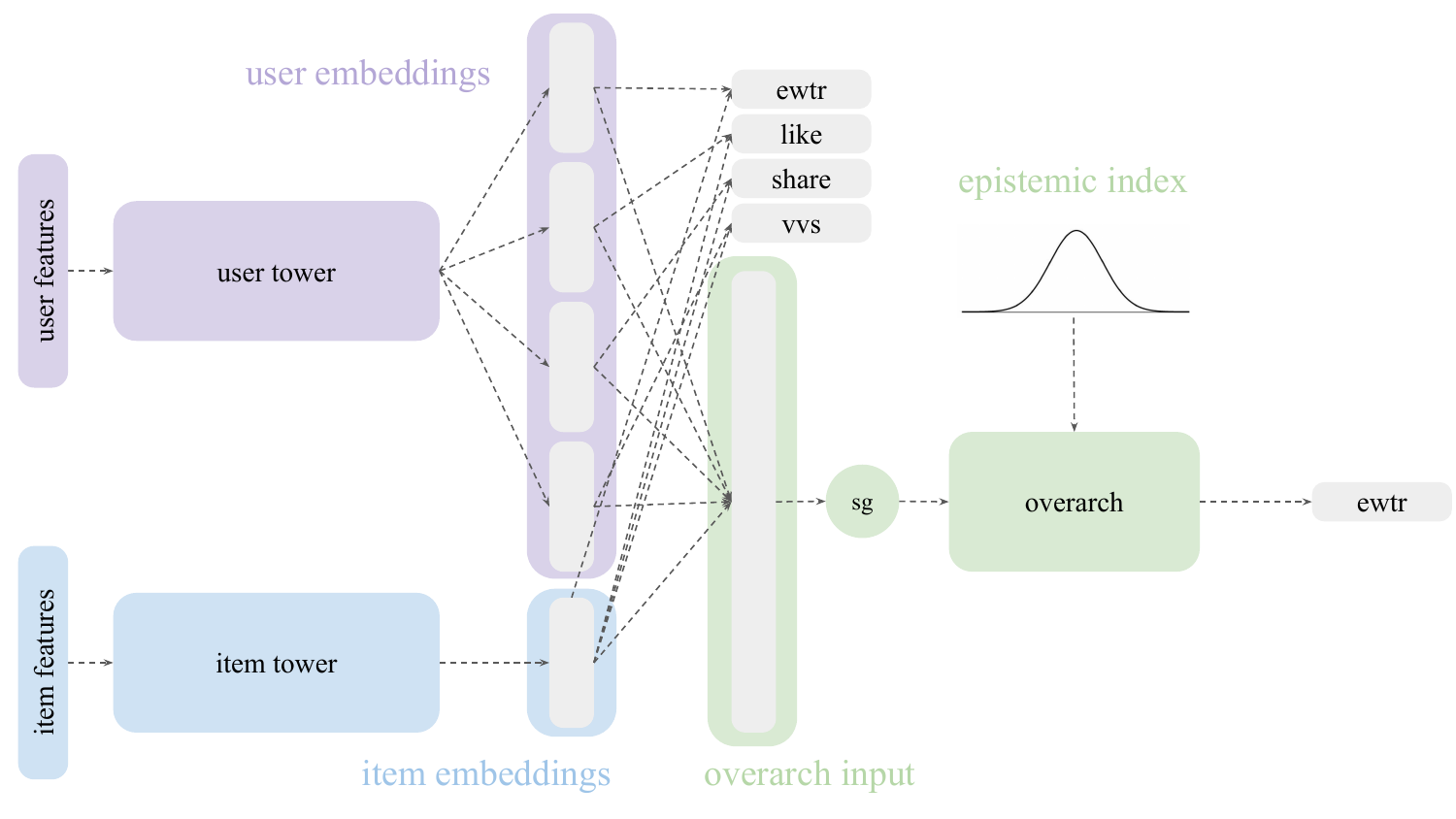}
    \caption{The above diagram depicts the training setup of our recommendation system.  Note that the epinet is part of the overarch component.}
    \label{fig:epinet_overarch}
\end{figure*}

Let $a\in \Ac$ denote an item selected by action $A_t$ and let $y_{t+1}$ be the labels associated with the outcome of showing item $a$ to user $\psi_t$.  We assume that $y_{t+1} \in [0,1]^K$ and we let $\tilde{y}_{t+1}\in [0,1]$ denote the single task label that we provide to the epinet.  We then sample a single epistemic indices $z_t$ from $\normal(0, I_{d_z})$.
\begin{align*}
    & \Lc(\xi_u, \xi_i, \theta, \eta, \phi_{t,a}, \psi_t, z_t)\\
    & = \underbrace{\sum_{k=1}^{K} {\rm BCE}\left(y_{t+1,k},\ g^{\rm (item)}_{\xi_i}(\phi_{t,a})^\top g^{\rm (user)}_{\xi_u}(\psi_t)_k\right)}_{\rm embedding\ loss}\\
    &\quad + \underbrace{{\rm BCE}\left(\tilde{y}_{t+1},\ f_{\theta,\eta}(x_t,z_t)\right)}_{\rm epinet\ loss},
\end{align*}

where ${\rm BCE}$ denotes binary cross entropy.  In practice, we sample a minibatch of actions and average the loss across the minibatch before taking a gradient step.

\subsection{Epinet Thompson Sampling Algorithm}

With the above details in place, we now present the epinet Thompson sampling algorithm.  

\begin{algorithm}[h]
	\caption{Epinet Thompson Sampling} 
	\begin{algorithmic}[1]
            \State Initialize trainable parameters $\xi^i_0, \xi^u_0, \theta_0, \eta_0$ and fixed parameters for prior net.
            \State Observe $O_0$
		\For {$t =0,1,2,\ldots$}
                \State Sample $z_t \sim \normal(0, I_{d_z})$
                \State Compute user embedding $g^{\rm (user)}_{\xi^u_t}(\psi_{t})$
                \For {$a = 1, 2, \ldots, N$}
                    \State Compute item embeddings $g^{\rm (item)}_{\xi^i_t}(\phi_{t,a})$
                    \State Compute overarch input $x_{t,a}$ with user and item embeddings
                \EndFor
                \State Select $M$ actions $a_1, \ldots, a_M \in \Ac$ which maximize 
                $f_{\theta_t, \eta_t}(x_{t,a}, z_t)$
                \State $A_t \leftarrow (a_1, a_2, \ldots, a_M)$
                \State Observe $O_{t+1}$
                \State $(\xi^u_{t+1}, \xi^i_{t+1}, \theta_{t+1}, \eta_{t+1}) \leftarrow (\xi^u_{t}, \xi^i_{t}, \theta_{t}, \eta_{t}) - \alpha \cdot \nabla\Lc(\xi^u_t, \xi^i_t, \theta_t, \eta_t, \phi_{t,a}, \psi_t, z_t, y_{t+1})$
		\EndFor
	\end{algorithmic} 
\end{algorithm}

Note that prior distribution is represented by the initial parameters of the various involved neural networks.  For epinet, this prior is a distribution over neural network \emph{functions}.  Sampling from this distribution involves first sampling an epistemic index $z_t$ from the reference distribution and then running a forward pass with $z_t$ held fixed.  Meanwhile, the approximate posterior update is performed via gradient descent on the objective.  In the following section, we outline the details of our online experimentation with epinet Thompson sampling for cold start retrieval.

\subsection{Recommendation Pipeline}

\begin{figure}[H]
    \centering
    \includegraphics[width=\linewidth]{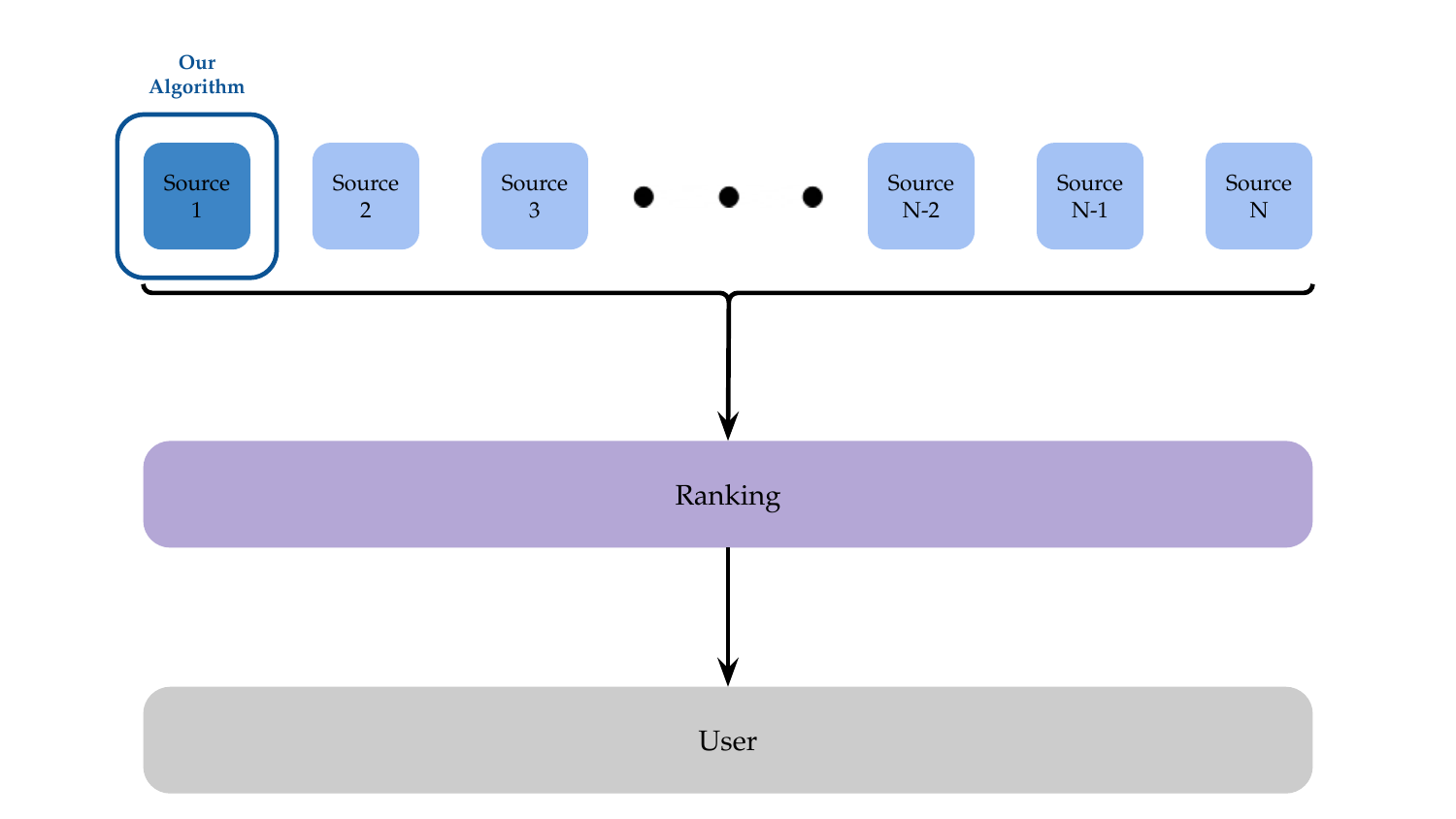}
    \caption{The above diagram depicts the content cold start funnel.  Each source filters a large batch of video content and passes the results to a blender before it is ranked.  Videos which achieve sufficient rank are presented to users and monitored for engagement. Our algorithm impacts the proposals of one of the sources which feed into the ranking algorithm.}
    \label{fig:ccs_funnel}
\end{figure}

The Facebook Reels recommendation system involves various phases in which recommendation proposals are generated, pooled, and then ranked before they are served to the user.  Since our project explores content cold start, our method operates at the proposal stage of recommendation.  Before the proposals are presented to the ranker, they are blended with the proposals of other sources which will be running their own separate recommendation algorithm.  Therefore, even if our algorithm suggests a piece of content to a particular user, it will not be served to that user unless the ranker predicts that our content is more promising than that of other sources.  However, quota is set aside for cold start content to mitigate popularity bias.  The application of epinets to the \emph{ranking} stage of recommendation remains an interesting direction for future research.

\section{Experiments}

We conducted an online A/B test over a period of 5 days to test the effectiveness of our proposed method.  For contextual bandit experiments, this is a standard duration as it doesn't take long for results to converge.  We run our test on Facebook's Reels recommendation system which serves billions of users each day.  Our method is applied at the retrieval stage of recommendation.  The generator which we deploy our experiment serves around $120$ million users per day.  For the test and control arms, we allocate groups of around $12$ million users.

\subsection{Implementation Details}
The user/item embeddings are trained to minimize loss on $4$ different tasks: watch score (ws), like, share, and video view seconds (vvs).  Watch score is defined below:
$${\rm ws} = \begin{cases}
    1 & \parbox[t]{6.5cm}{if video length $< 10$ seconds and user completed video more than once}\\
    1 & \parbox[t]{6.5cm}{ if video length $\geq 10$ seconds and $< 20$ seconds and user completed video}\\
    1 & \parbox[t]{6.5cm}{ if video length $\geq 20$ seconds and user watched at least $20$ seconds}\\
    0 & \text{ otherwise}
\end{cases}.$$
Like and share are a binary signals which are $1$ if the user liked/shared the video and $0$ otherwise.  Vvs is defined below:
$${\rm vvs} = \begin{cases}
    0 & \text{ if video viewed for } < 10 \text{ seconds}\\
    \frac{1}{9} & \text{ if video viewed for } \geq 10 \text{ seconds and } < 20 \text{ seconds}\\
    \vdots & \vdots\\
    1 & \text{ if video viewed for } \geq 90 \text{ seconds}\\
\end{cases}.$$

While the user and item embeddings are trained on the $4$ above labels, the overarch is only trained on ws.  The control arm only maintains a point estimate and does \emph{not} include an overarch component.  Therefore, it is trained to minimize BCE loss on the above signals and recommends content greedily with respect to its current point estimate.  Both models are trained every hour, initialized from the most recent checkpoint.  They are trained on a pool of new data which is aggregated across \emph{all} other generators.  Therefore, the data contributed by our algorithm to the aggregated pool is very miniscule.  This brings about two concerns $1)$ data leakage between treatment arms, $2)$ dilution of data collected by our method by data from other generators.  However, upon inspection the embeddings for each user were sufficiently different.  As a result, there is $1)$ no meaningful generalization to users \emph{across} treatment groups and $2)$ less concern of dilution since each the data from other generators is sufficiently disjoint in embedding space.  We hypothesize that theses are the reasons that we observe significant performance improvements with our method despite the above concerns.

Each embedding is of dimension $d=128$ and the epistemic index dimension $d_z = 5$.  In the overarch model, we use 2-hidden layer MLPs with hidden dimensions $[384, 256]$ for both the epinet and the base mlp with glorot initialization.  As aforementioned, we sample only a single epistemic index for both training and inference. 

\subsection{Experimental Results}
We now outline the results of our online experiments.  A recommendation counts as an ``impression'' if it is displayed on the user's screen for at least $250$ milliseconds.  Since we are interested in the performance of cold start content, we measure performance on content with fewer than 10000 impressions.  We further stratify performance 
by grouping the statistics by video impression count \emph{buckets} of $[0:100, 100:200, 200:400, 400:1000, 1000:2000, 2000:3000, 3000:4000, 4000:5000, 5000:10000]$.  We report improvements across $3$ efficiency metrics: like per impression, video completion per impression, and watch score per impression.  Video completion is $1$ if the user viewed the video to completion and $0$ otherwise.

\begin{figure}[H]
    \centering
    \includegraphics[width=\linewidth]{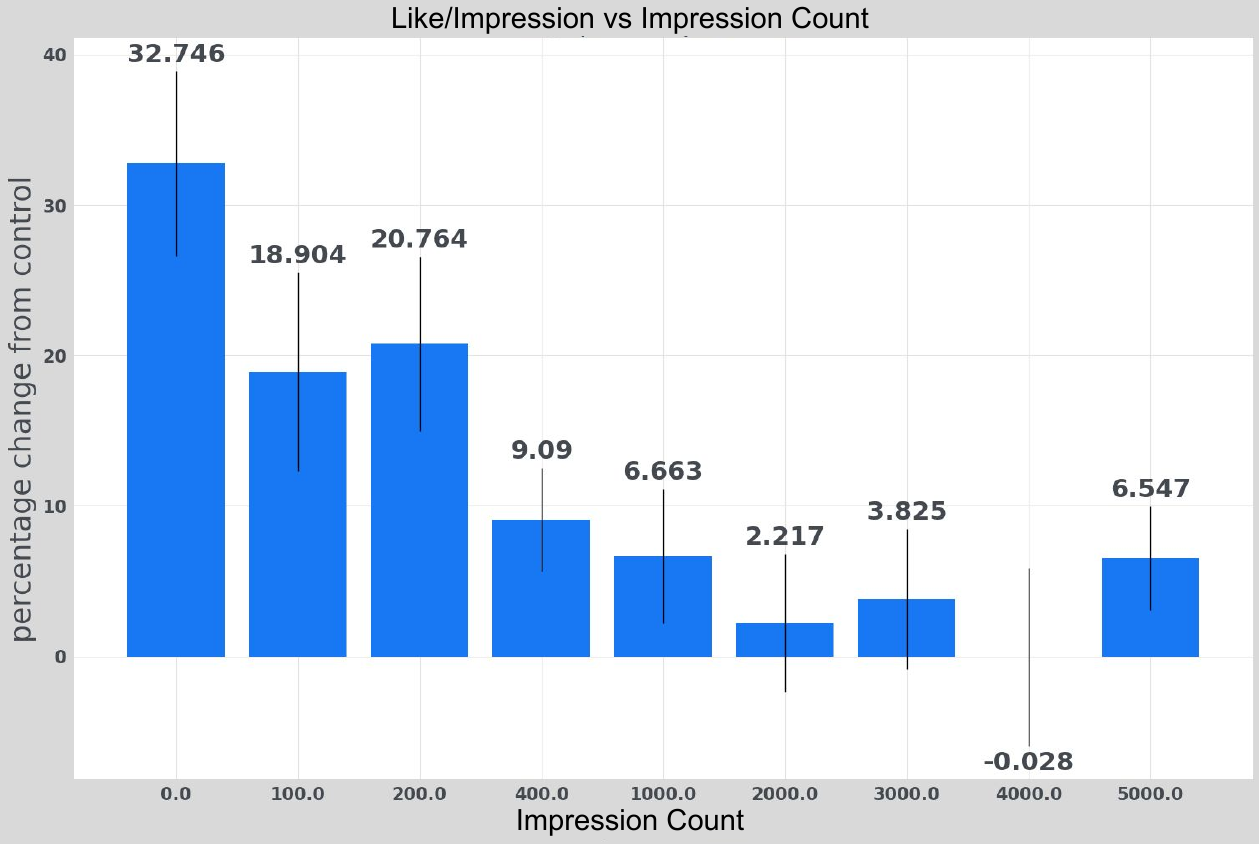}
    \caption{We depict percentage change in like per impression between our method and the control.  We group videos by their impression counts.  We notice the most significant benefits in the videos with lower impression count which is promising for content cold start.}
    \label{fig:like}
\end{figure}

\begin{figure}[H]
    \centering
    \includegraphics[width=\linewidth]{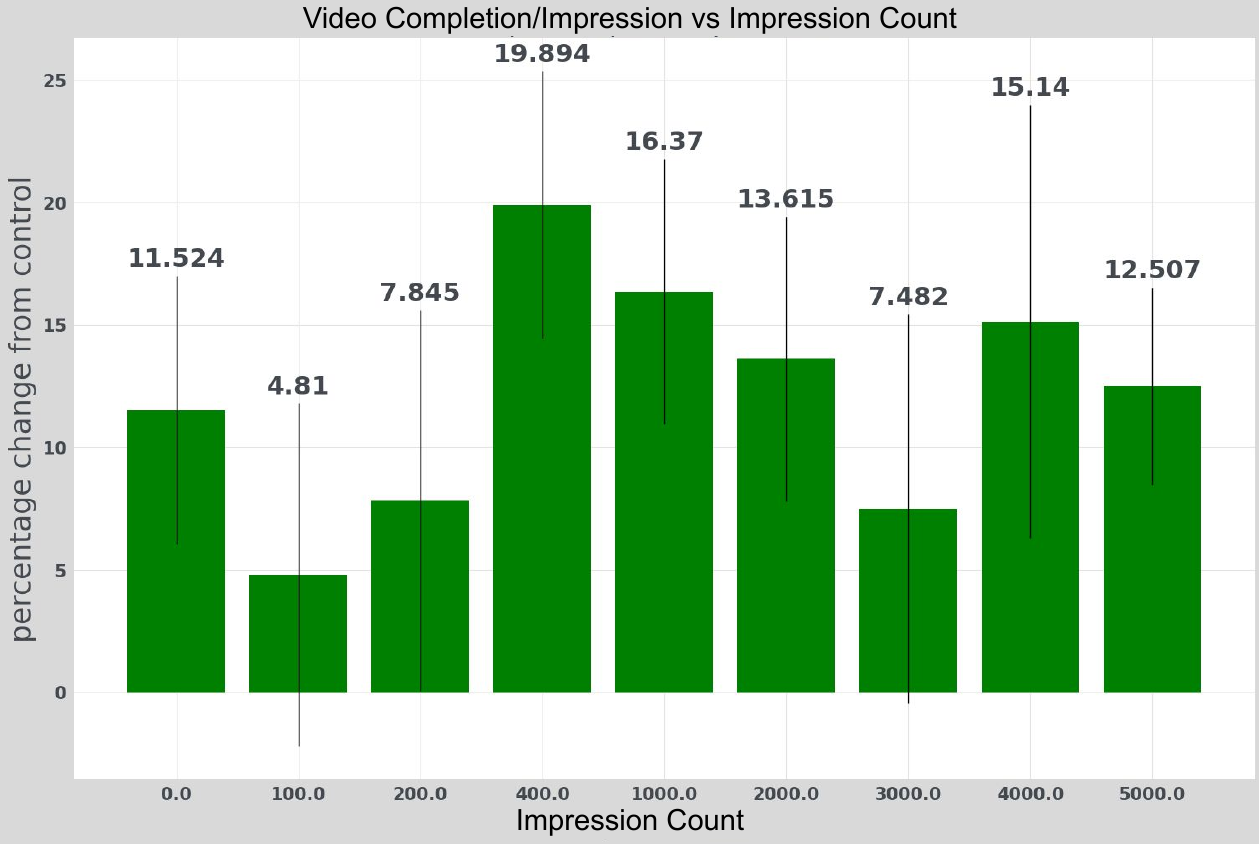}
    \caption{We depict percentage change in video view completion per impression between our method and the control.  We group videos by their impression count.  We notice improved video view completion across all impression counts. }
    \label{fig:vvp100}
\end{figure}

\begin{figure}[H]
    \centering
    \includegraphics[width=\linewidth]{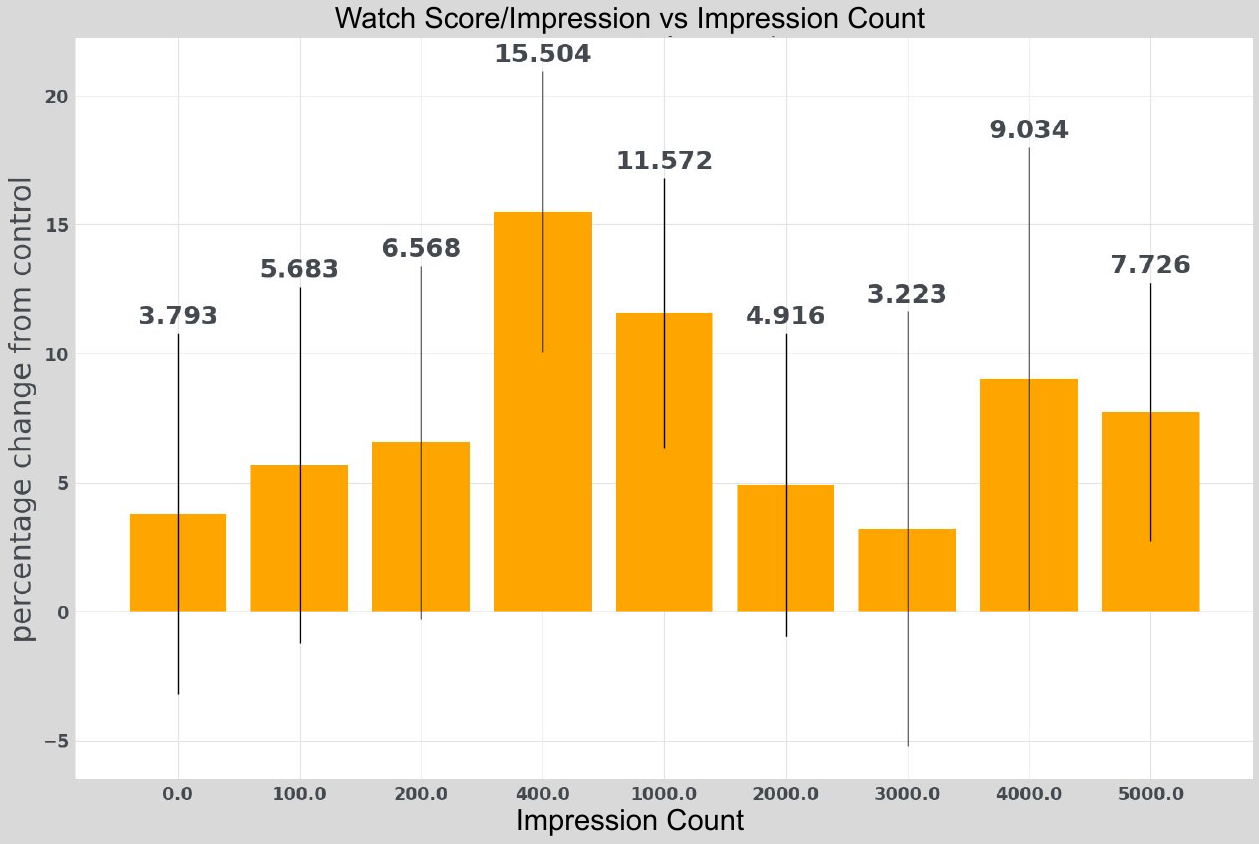}
    \caption{We depict percentage change in watch score per impression between our method and the control.  We group videos by their impression count.}
    \label{fig:ewtr}
\end{figure}

Figures \ref{fig:like}, \ref{fig:vvp100}, and \ref{fig:ewtr} depict the percentage change of our epinet algorithm in comparison to the control group for the various aforementioned efficency metrics.  We plot the $95\%$ confidence intervals for each metric and impression count bucket.  Hence, a change is significant if the error bar does not cross past $0$.   
 Figure \ref{fig:like} demonstrates large improvements in like per impression, especially for content with lower impression counts.  This suggests that via intelligent exploration, the system can more reliably explore cold start content and serve videos that the users enjoy.  Figure \ref{fig:vvp100} demonstrates significant improvement in video view completion per impression, indicating that the epinet algorithm is successfully able to recommend content which users enjoy enough to watch to completion.  Finally, Figure \ref{fig:ewtr} depicts positive, though less significant, results for watch score per impression.

\begin{figure}[H]
    \centering
    \includegraphics[width=\linewidth]{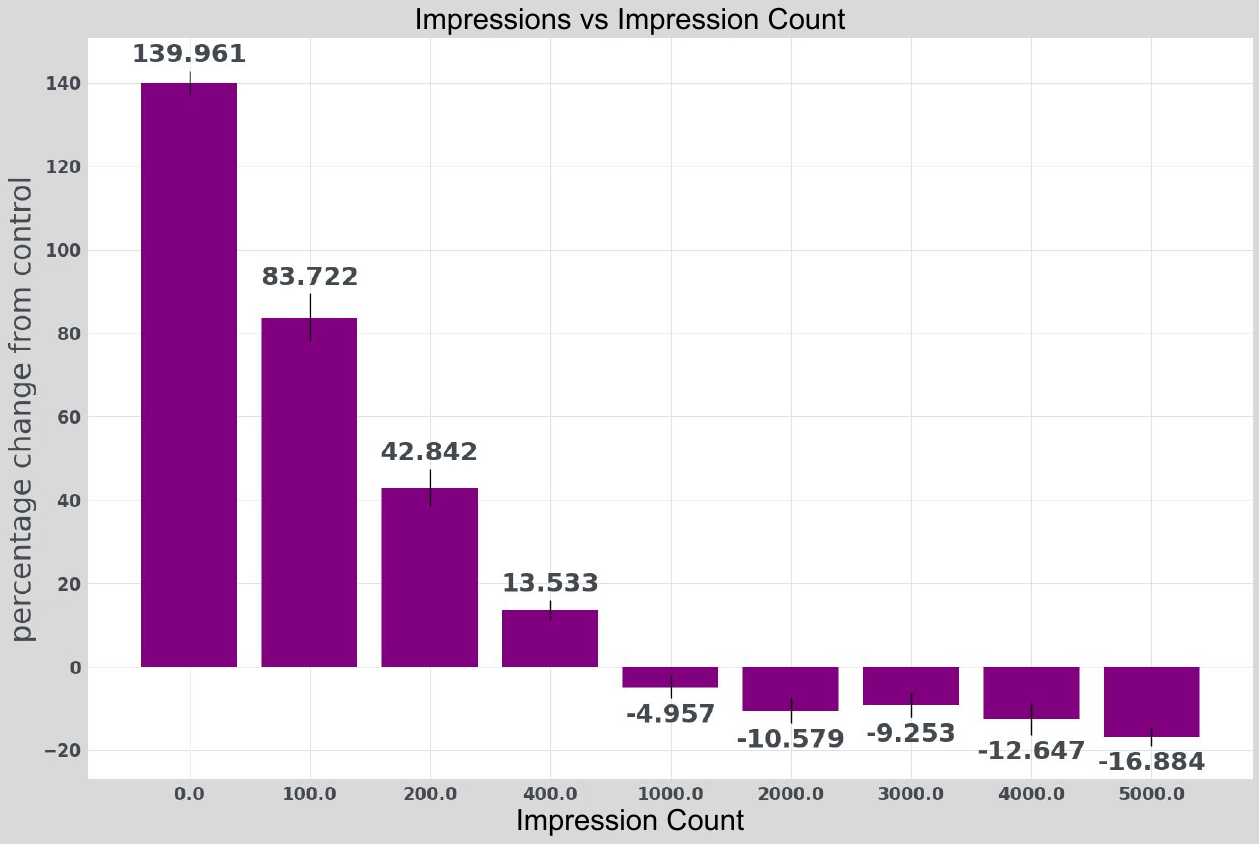}
    \caption{We depict the percentage change in impressions for content grouped by the impression count for the content.  Our method dramatically improves impressions for videos with fewer impressions and modestly pulls traffic from videos with higher impression count.}
    \label{fig:vpv}
\end{figure}

Finally, we analyze how our method reallocates recommendations across the impression count buckets.  Figure \ref{fig:vpv} shows that our method dramatically boosts impressions for content with fewer impressions and modestly pulls traffic from content with more impressions.  We also note that our method demonstrates an overall $17\%$ boost in impressions when aggregated across content of all impression counts. 

The results of this empirical investigation suggest that applying epinets to approximate Thompson sampling can provide concrete improvements in an online production system.  

\section{Conclusion}
Our work marks the first empirical investigation of epinets in an online production recommendation system.  The investigation demonstrated that epinets can provide an effective solution to managing the exploration-exploitation trade-off in content cold start.  Interesting future extensions of this work include larger-scale experiments and application of these ideas to later stages (ranking) in the recommendation pipeline.  Extensions from the contextual bandit to a reinforcement learning setting also serves as an interesting direction for future research.


\bibliographystyle{ACM-Reference-Format}
\bibliography{sample-base}

\appendix









\end{document}